\documentclass[pdflatex,sn-nature]{sn-jnl}

\usepackage{graphicx}%
\usepackage{multirow}%
\usepackage{amsmath,amssymb,amsfonts}%
\usepackage{amsthm}%
\usepackage{mathrsfs}%
\usepackage[title]{appendix}%
\usepackage{xcolor}%
\usepackage{textcomp}%
\usepackage{manyfoot}%
\usepackage{booktabs}%
\usepackage{algorithm}%
\usepackage{algorithmicx}%
\usepackage{algpseudocode}%
\usepackage{listings}%

\usepackage{scalefnt}
\AtBeginDocument{\scalefont{1.2}}
\usepackage{caption}
\usepackage{setspace}
\begin{document}

\title[Article Title]{\huge Facet-selective ballistic supercurrent in a weak topological insulator}

\author*[1,2]{Prasanna Rout}
\email{prasanna.rout@physics.gu.se}
\equalcont{These authors contributed equally to this work.}

\author[1]{Ankit Khola}
\equalcont{These authors contributed equally to this work.}
\author[1]{Lalit Pandey}
\author[3]{Paolo Sessi}
\author[4]{Xiaochun Huang}
\author[1]{Ivo Cools}
\author[5]{S. Galeski}
\author[4]{Matthias Bode}
\author[2,6,7]{Johan Åkerman}
\author[1,8]{Floriana Lombardi}
\author[1]{Thilo Bauch}
\author*[1,8]{Saroj P. Dash}
\email{saroj.dash@chalmers.se}

\affil[1]{Department of Microtechnology and Nanoscience, Chalmers University of Technology, SE-41296, Göteborg, Sweden}

\affil[2]{Department of Physics, University of Gothenburg, Göteborg 41296, Sweden}

\affil[3]{Max-Planck-Institut für Festkörperforschung, Heisenbergstraße 1, 70569, Stuttgart, Germany}

\affil[4]{Physikalisches Institut, Experimentelle Physik II, Universität Würzburg, Am Hubland, 97074 Würzburg, Germany}

\affil[5]{Institute for Molecules and Materials, Radboud University, Heyendaalseweg 135, 6525 AJ Nijmegen, The Netherlands}

\affil[6]{Center for Science and Innovation in Spintronics, Tohoku University, 2-1-1 Katahira, Aoba-ku, Sendai, 980-8577, Japan}

\affil[7]{Research Institute of Electrical Communication, Tohoku University, 2-1-1 Katahira, Aoba-ku, Sendai, 980-8577, Japan}

\affil[8]{Wallenberg Initiative Materials Science for Sustainability, Department of Microtechnology and Nanoscience, Chalmers University of Technology, SE-41296, Göteborg, Sweden}

\renewcommand{\abstractname}{}
\abstract{\large \setstretch{1} Topological superconductivity is widely pursued by inducing superconducting correlations in topologically protected boundary states. In two dimensions, this strategy has been realized using one‑dimensional topological edge modes, but in three‑dimensional crystals, spatially separated surface supercurrents confined to selected facets have not yet been achieved. Here we demonstrate facet-selective ballistic supercurrent in Josephson junctions based on the weak topological insulator ZrTe$_5$. Superconducting quantum interferometry reveals SQUID-like critical current oscillations with flux-quantum periodicity, establishing that the supercurrent is spatially concentrated on specific crystallographic facets that host gapless topological surface states. Rotating the magnetic field yields markedly distinct interference patterns, linking the supercurrent distribution to the underlying bulk topology. The exponential temperature dependence of the critical current and triangular interference lobes provide signatures of ballistic transport due to high-transmission topological channels. These results establish weak topological insulators as a platform for facet-resolved superconducting devices and higher‑order topological superconductivity.}




\maketitle
Engineering a topological superconducting state that hosts non-Abelian Majorana quasiparticles has been proposed as a basis for topologically protected qubits~\cite{Qi_Topological_2011,sato2017topological}. Because intrinsic topological superconductors are rare, a practical route is to proximitize a topological material with a conventional superconductor so that Cooper pairs leak into the topologically protected boundary states, thereby converting these states into topological superconducting channels. In two-dimensional topological systems, Josephson supercurrents have been demonstrated in the helical edge channels of quantum spin Hall insulators and in chiral edge modes of quantum Hall systems, where the supercurrent is confined to one-dimensional topological edge channels~\cite{hart_induced_2014,pribiag2015edge,amet_supercurrent_2016,vignaud_evidence_2023,rout_supercurrent_2024}. 

\begin{figure}[t]
\centering
\includegraphics[width=\textwidth]{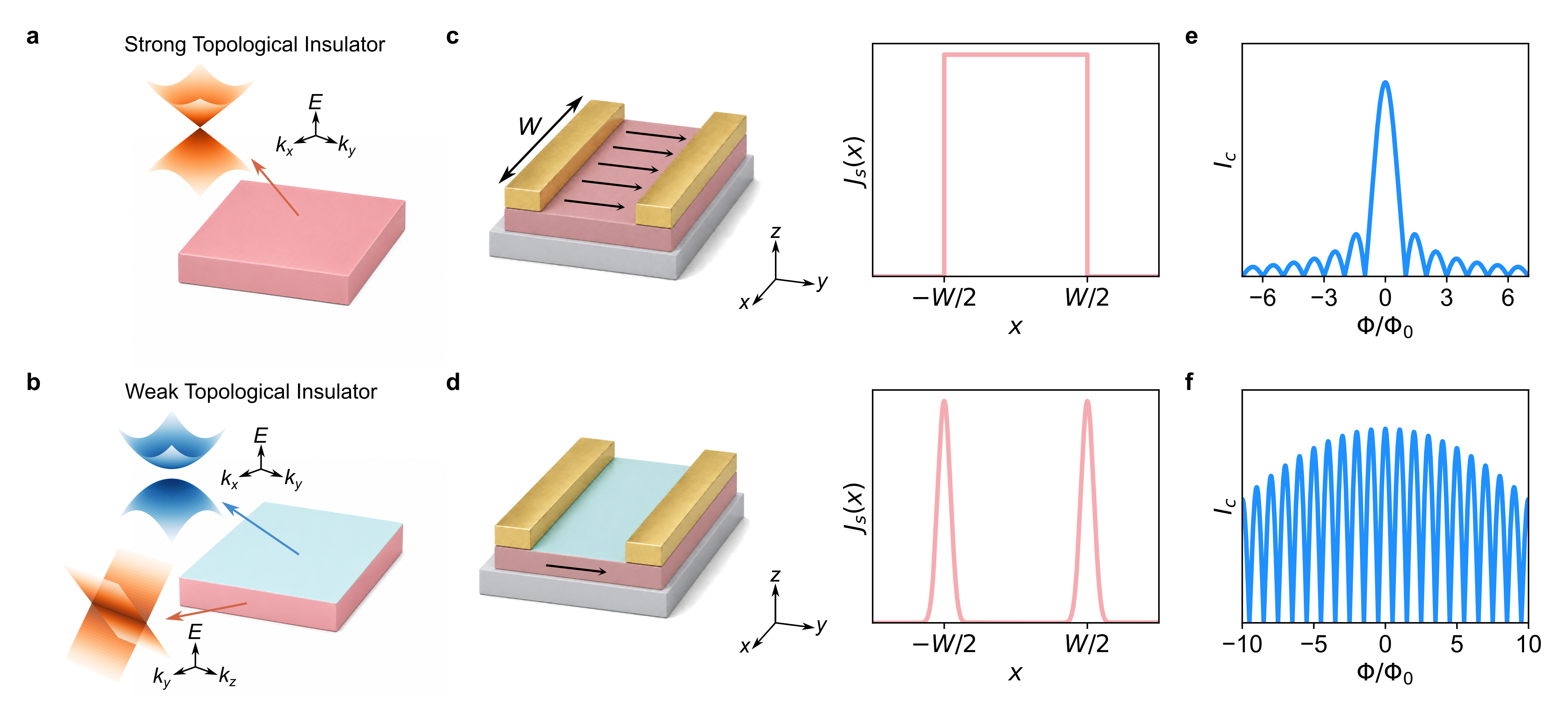}
\caption{\textbf{Probing the boundary states in three-dimensional topological insulators by superconducting quantum interferometry.} \textbf{a, b}, Schematic band structure of a strong topological insulator (STI, \textbf{a}), and a weak topological insulator (WTI, \textbf{b}). An STI hosts gapless Dirac cones on the time-reversal-symmetric crystal facets (pink surfaces) while the bulk is insulating. In a WTI, conducting surface states due to one-dimensional Dirac lines reside  only on specific crystallographic facets such as $y$–$z$ and $z$–$x$ (pink surfaces), whereas other $x$–$y$ facets (blue surfaces) remain fully gapped. \textbf{c, d}, Josephson junctions based on an STI and a WTI coupled to superconducting electrodes. In the STI junction (\textbf{c}), the supercurrent flows homogeneously, producing a rectangular $J_s(x)$ profile across the junction width $W$. In contrast, the supercurrent in the WTI junction (\textbf{d}) is confined to two side facets, as only $y$–$z$ surfaces are conducting. \textbf{e, f}, Expected superconducting quantum interference patterns, i.e., the critical current $I_c$ as a function of perpendicular magnetic field $B_z$ or magnetic flux $\Phi$, corresponding to the junction types shown in (\textbf{c,d}). The STI junction exhibits a Fraunhofer pattern, whereas the WTI junction displays a SQUID-like interference.}\label{Fig-schem}
\end{figure}

Extending this physics to three-dimensional materials is a natural and important step, in which one-dimensional edges are now replaced by two-dimensional surface states on the crystallographic facets of a bulk crystal. The bulk topological class of a three-dimensional topological insulator determines which crystal facets support gapless surface states~\cite{fu2007topological,Hasan_Topological_2010,Qi_Topological_2011}. In a strong topological insulator (STI), gapless surface states reside on every crystal facet (Fig.~\ref{Fig-schem}\textbf{a}). When an STI is coupled to a superconductor, superconducting correlations therefore extend over the entire junction area, and superconducting quantum interferometry produces the conventional Fraunhofer diffraction pattern of a homogeneous junction (Fig.~\ref{Fig-schem}\textbf{c,e}). Josephson junctions based on STIs have demonstrated proximity-induced superconductivity in topological surface states~\cite{sacepe2011gate,veldhorst2012josephson,charpentier2017induced}. In contrast, weak topological insulators (WTIs), which can be viewed as layered stacks of quantum spin Hall insulators, support gapless topological surface states only on selected facets, while the remaining surfaces are gapped (Fig.~\ref{Fig-schem}\textbf{b})~\cite{ringel2012strong,lau2015one,noguchi2019weak,zhong2023towards,yu2024observation,xu2024realization,zhang2017electronic,mutch2019evidence,chen2025first,wu2016evidence,li2016experimental,zhang2021observation}. This intrinsic facet selectivity implies that the supercurrent in a WTI-based Josephson junction predominantly flows through two spatially separated surface channels associated with the conducting side facets (Fig.~\ref{Fig-schem}\textbf{d}). These two parallel supercurrent paths can make the junction behave as an effective SQUID, producing an interference pattern with periodic critical current oscillations as a function of magnetic field (Fig.~\ref{Fig-schem}\textbf{f}). Such facet‑selective WTI junctions may provide an experimental setting for exploring higher‑order topological superconductivity, where topological hinge and corner modes are predicted to appear~\cite{luo2021higher,schindler2018higher_sci,schindler2018higher,choi2020evidence,kononov2020one,li2020reducing}. Despite these expectations, Josephson coupling in a WTI has not been realized, owing to the scarcity of well‑characterized materials and the difficulty of disentangling boundary‑state transport from contributions due to bulk states, trivial surface states, and disorder~\cite{ando2013topological,zhong_coalescence_2025,zhang2021observation,kovacs2023revealing}.

Here we demonstrate, for the first time, facet-selective ballistic supercurrent in the three-dimensional weak topological insulator ZrTe$_5$. Using superconducting quantum interferometry in planar Josephson junctions, we show that the supercurrent is predominantly confined to two spatially separated side facets, where gapless topological surface states are expected. This facet selectivity is robust across multiple devices and two different superconductors (Al and NbN), and its topological origin is further supported by a striking dependence on magnetic-field orientation. The ballistic, high-transmission nature of the transport channels is independently revealed by the temperature dependence of the critical current and the shape of the interference pattern. Together, these observations show that the bulk topology of a three-dimensional crystal can directly dictate where supercurrent flows, enabling the possibility of facet-engineered topological superconducting devices.

\begin{figure}[t]
\centering
\includegraphics[width=1\textwidth]{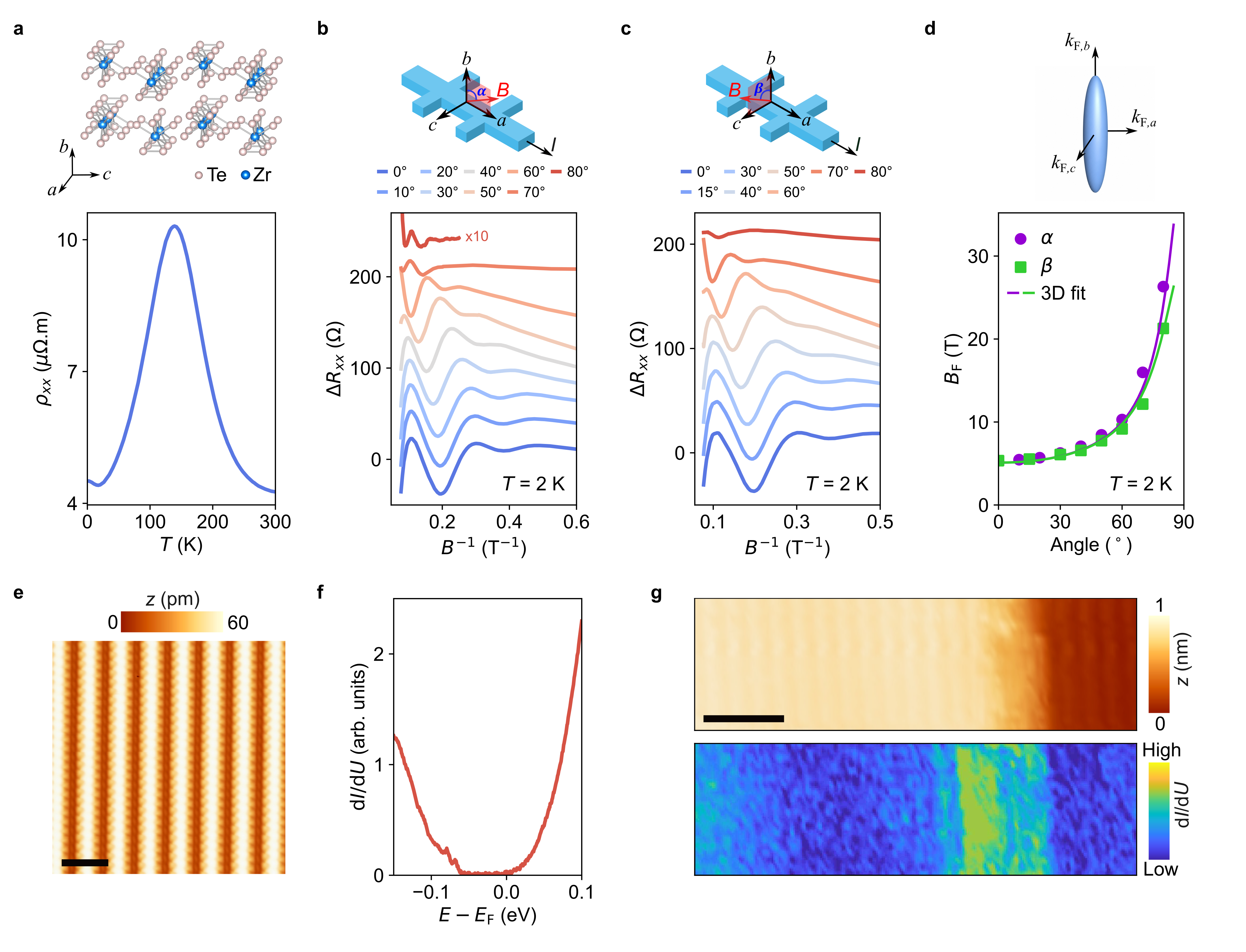}
\caption{\textbf{Electronic transport and spectroscopic measurements of ZrTe$_5$.} \textbf{a}, Temperature dependence of the longitudinal resistivity $\rho_{xx}$ measured with current applied along the \textit{a}-axis. The top panel shows the crystal structure of ZrTe$_5$, consisting of two-dimensional layers in the $a$–$c$ plane that are coupled by van der Waals forces along the $b$-axis. In the WTI regime, the $a$–$c$ surfaces are gapped, while topological surface states reside on the $a$–$b$ and $b$–$c$ facets. \textbf{b, c}, Magnetoresistance $\Delta R_{xx}(B)$ after background subtraction, measured at 2~K for various field angles $\alpha$ in the \textit{a–b} plane and $\beta$ in the \textit{b–c} plane. Here, $\alpha$ and $\beta$ denote the angle between the \textit{b}-axis and the magnetic field as illustrated schematically in the top panels. \textbf{d}, Angular dependence of $B_{\rm{F}}$ for field rotations within the \textit{a–b} and \textit{b–c} planes. The data are fitted using the expression for a three-dimensional ellipsoidal Fermi surface (solid lines). The top panel shows the reconstructed Fermi surface with Fermi wave vectors $k_{\rm{F},\mathit{a}}$, $k_{\rm{F},\mathit{b}}$ and $k_{\rm{F},\mathit{c}}$ along the \textit{a}-, \textit{b}- and \textit{c}-crystallographic directions, respectively. \textbf{e}, STM image showing the topography of the surface exposed after cleaving. Scale bar is 2 nm. \textbf{f}, Differential conductance ${\rm d}I/{\rm d}U$ as a function of energy relative to the Fermi level $E_{\rm{F}}$.  \textbf{g}, Topography (top panel) and corresponding ${\rm d}I/{\rm d}U$ map (bottom panel) measured across a step edge. The ${\rm d}I/{\rm d}U$ map represents the LDOS distribution at an energy $E - E_{\rm{F}} = -31$~meV (i.e. inside the bulk gap), showing the spatial distribution of the electronic states. Scale bar is 5 nm.}\label{Fig-transport}
\end{figure}

\section*{Results}
\subsection*{Transport and spectroscopic signatures of weak topological insulator}

To determine whether ZrTe$_5$ is in the WTI or STI phase, we investigate the electronic transport and local spectroscopic properties of ZrTe$_5$. The longitudinal resistivity of a ZrTe$_5$ Hall bar device exhibits a pronounced peak at $T_p \approx 139$ K (Fig.~\ref{Fig-transport}\textbf{a}), which is attributed to a Lifshitz transition associated with a crossover from a hole-type insulating state to an electron-type metallic state upon cooling through $T_p$ \cite{zhang2017electronic,galeski2021origin,tang2019three,kovacs2023revealing}. This transition is further confirmed by the Hall measurements presented in Supplementary Fig.~S1. We observe clear Shubnikov–de Haas (SdH) oscillations in the longitudinal resistance $R_{xx}(B)$ in two different field configurations: one with the magnetic field rotated within the \textit{a–b} plane (Fig.~\ref{Fig-transport}\textbf{b}) and the other within the \textit{b–c} plane (Fig.~\ref{Fig-transport}\textbf{c}). A single SdH frequency $B_{\rm{F}}$ is observed for both field orientations, indicating a single Fermi pocket associated with an electron-like band~\cite{galeski2021origin,tang2019three}. The angular dependence of $B_{\rm{F}}$ presented in Fig.~\ref{Fig-transport}\textbf{d} is well described by a three-dimensional ellipsoidal Fermi surface with $k_{\rm{F},\mathit{a}(\mathit{c})} \ll k_{\rm{F},\mathit{b}}$ (see Supplementary Note 3). To probe the bulk gap and surface states, we performed scanning tunnelling microscopy and spectroscopy measurements on ZrTe$_5$ crystals cleaved in ultrahigh vacuum. Fig.~\ref{Fig-transport}\textbf{e} presents an atomically resolved topographic image revealing a quasi-one-dimensional modulation arising from the alternating ZrTe$_3$ chains (see the crystal structure in Fig.~\ref{Fig-transport}\textbf{a}). This observation confirms that the exposed surface is the \textit{a–c} facet. The differential conductance ${\rm d}I/{\rm d}U$, which provides information about the local density of states (LDOS), reveals an energy gap of $\approx 70$~meV (Fig.~\ref{Fig-transport}\textbf{f}). Moreover, we spectroscopically map the LDOS across a step edge. Fig.~\ref{Fig-transport}\textbf{g} shows a topographic image revealing two terraces separated by a step with a height corresponding to the separation between adjacent ZrTe$_5$ layers (0.8 nm). The corresponding ${\rm d}I/{\rm d}U$ map reveals a clear enhancement of the LDOS along the step edge, signaling the presence of conducting states (see Supplementary Note 4 for additional spectroscopic data). Notably, these states extend into the terrace over a spatial width of $\approx 6$~nm and remain continuously distributed along the edge. The coexistence of a bulk gap on the \textit{a–c} facet and spatially extended conducting channels along the step edge provides strong evidence that ZrTe$_5$ is in the WTI phase~\cite{li2016experimental,wu2016evidence}.

\subsection*{Proximity-induced superconductivity in ZrTe$_5$}

\begin{figure}[t]
\centering
\includegraphics[width=1\textwidth]{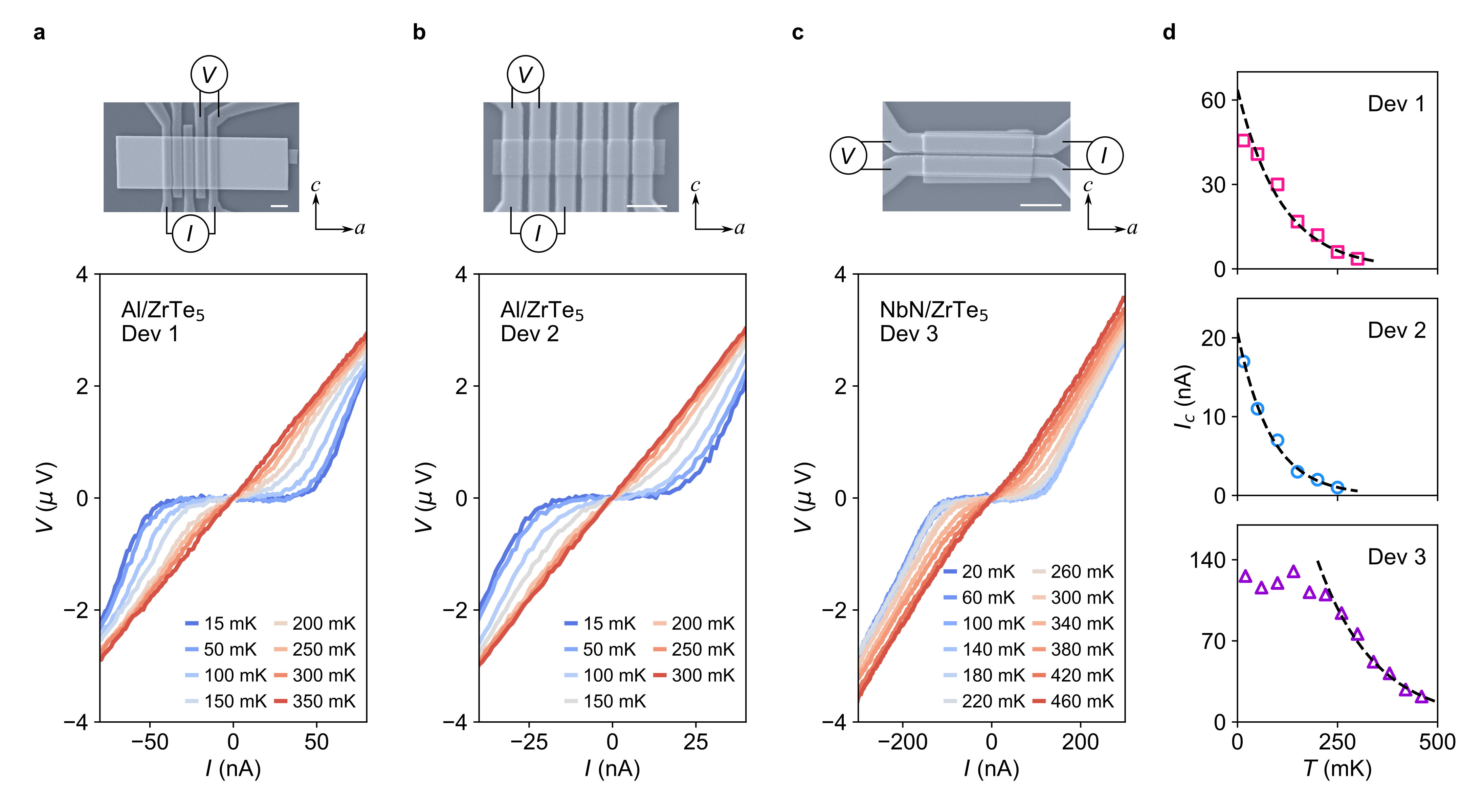}
\caption{\textbf{Supercurrent transport in ZrTe$_5$ Josephson junctions.} 
\textbf{a,b}, Voltage $V$ across the Al/ZrTe$_5$ junction as a function of applied current $I$ flowing along the \textit{a}-axis, measured at different temperatures. The data for Dev~1 and Dev~2 are presented in \textbf{a} and \textbf{b}, respectively. \textbf{c}, $V$($I$) curves for the NbN/ZrTe$_5$ junction (Dev~3), where the current is along the \textit{c}-axis. The top panels in \textbf{a}-\textbf{c} show scanning electron microscopy images of the junctions with the measurement schematics (Scale bar is 2~$\mu$m). \textbf{d}, Temperature dependence of $I_c$ for three junctions, extracted from \textbf{a}-\textbf{c}, together with fits (dashed lines) to the data  in the high‑temperature regime.}\label{Fig-inducedSC}
\end{figure}

We fabricated Josephson junctions by coupling ZrTe$_5$ to two different superconductors, Al and NbN, with junctions oriented along the \textit{a}- and \textit{c}-axes. The relatively long superconducting coherence length of Al is advantageous for establishing induced superconductivity in long junctions. In contrast, the higher critical field of NbN permits measurable critical currents in the presence of high magnetic fields, which will be discussed later. Fig.~\ref{Fig-inducedSC}\textbf{a,b} present the temperature dependence of current–voltage characteristics of two Al/ZrTe$_5$ junctions, while Fig.~\ref{Fig-inducedSC}\textbf{c} shows the corresponding data for an NbN/ZrTe$_5$ junction. The observation of a well-defined critical current $I_c$ demonstrates proximity-induced superconductivity in ZrTe$_5$ regardless of superconductor choice or crystal orientation.

Having confirmed induced superconductivity in all junctions, we now examine the temperature dependence of $I_c$ (Fig.~\ref{Fig-inducedSC}\textbf{d}). In the long ballistic junction limit, the discrete level spacing $\delta E$ of the Andreev bound states sets the relevant energy scale, and the supercurrent therefore follows a thermally activated dependence, $I_c \propto \exp\left(-2 \pi ^2 k_{\rm{B}} T / \delta E\right)$, where $k_{\rm{B}}$ is the Boltzmann constant~\cite{kulik1970macroscopic,bardeen1972josephson,Borzenets2016Ballistic}. All three junctions exhibit this exponential temperature dependence. The observed ballistic behavior strongly suggests that the supercurrent is predominantly carried by topological surface states rather than bulk channels. In a purely bulk scenario, the estimated mean free path is too short to support ballistic transport over the junction length (Supplementary Note~2), whereas topological surface states with suppressed backscattering can naturally sustain such ballistic supercurrents.


\begin{figure}[t]
\centering
\includegraphics[width=0.9\textwidth]{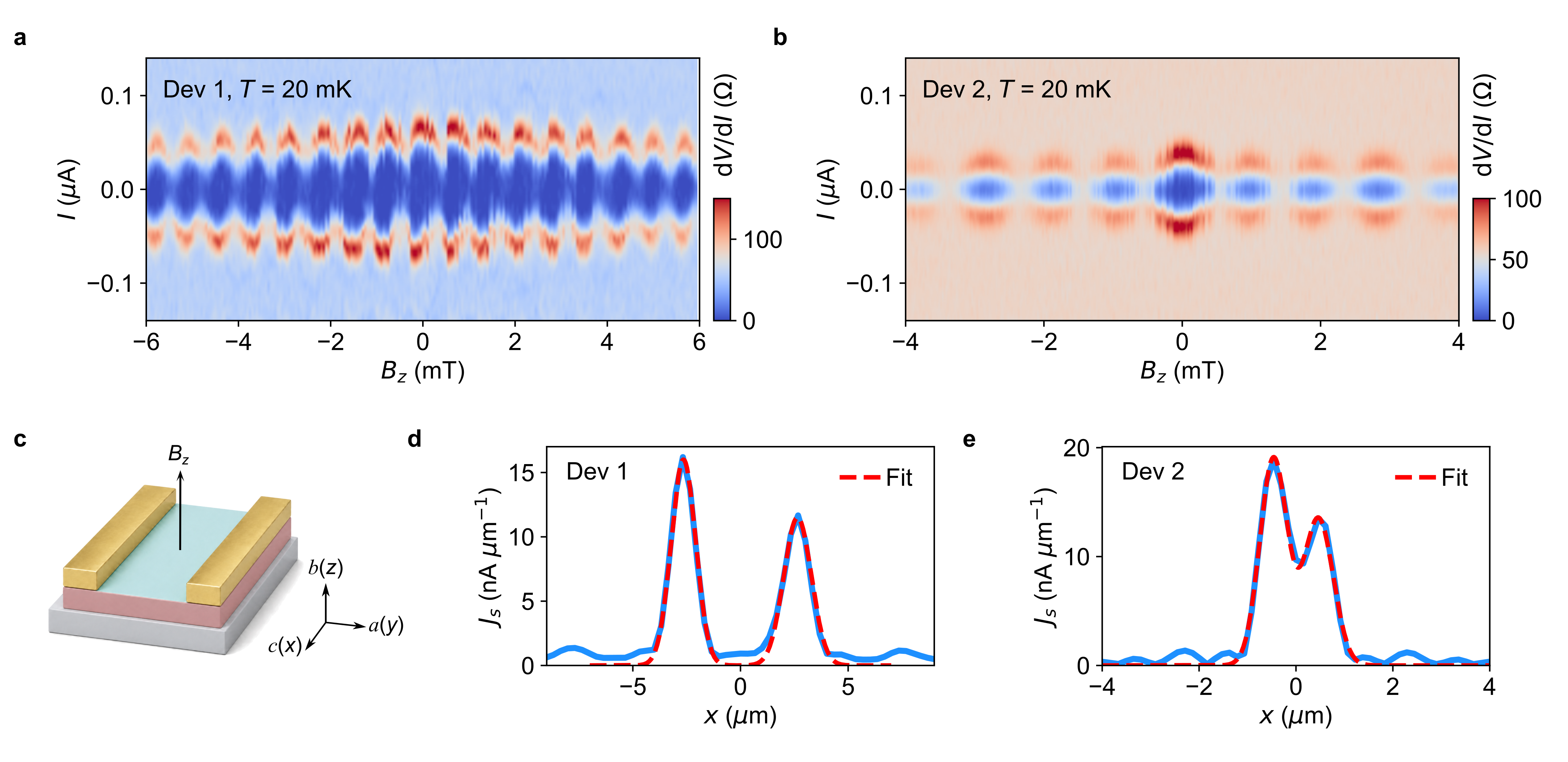}
\caption{\textbf{Superconducting quantum interferometry revealing spatially localized supercurrent in Al/ZrTe$_5$ junctions.} Superconducting quantum interference patterns for Dev~1 (\textbf{a}), and Dev~2 (\textbf{b}) measured at 20~mK, exhibiting clear SQUID-like oscillations. \textbf{c}, Schematics of the junctions, where the current flows along the \textit{a}-axis and the magnetic field is applied along the \textit{b}-axis. \textbf{d,e}, Corresponding supercurrent density profiles $J_s(x)$ extracted from the interference patterns in \textbf{a,b}. Two pronounced peaks indicate that the supercurrent is localized near opposite $a$–$b$ facets. The spatial supercurrent distributions are fitted by a combination of two Gaussian peaks (dashed lines).}\label{Fig-edgeSC}
\end{figure}

\subsection*{Facet-selective supercurrent probed by superconducting quantum interferometry}

The central evidence for boundary-state-mediated superconductivity comes from superconducting quantum interference measurements. For an STI or trivial junction with a uniform current distribution, the interference pattern is a Fraunhofer pattern (Fig.~\ref{Fig-schem}\textbf{c,e}). In contrast, a WTI-based junction with a unique surface-localized supercurrent is expected to produce SQUID oscillations (Fig.~\ref{Fig-schem}\textbf{d,f}). Fig.~\ref{Fig-edgeSC}\textbf{a,b} present the interference patterns of two Al/ZrTe$_5$ junctions, exhibiting equally spaced critical current oscillations whose amplitude decays slowly with increasing magnetic field. Quantitative analysis of the minima positions confirms integer flux quantum spacing of these oscillations (Supplementary Note 5), further demonstrating that the patterns are SQUID-like rather than Fraunhofer-like. Furthermore, we extract the supercurrent density profile $J_s(x)$ across the junction width from the interference patterns (see Supplementary Note 6 for details). The reconstructed profiles show two pronounced peaks near the junction edges, with Gaussian half-widths of 0.7-1.5~$\mu$m (Fig.~\ref{Fig-edgeSC}\textbf{d,e}), which are substantially larger than the conducting channel width observed by STM (Fig.~\ref{Fig-transport}\textbf{g}). Such broadening may reflect finite penetration of boundary states into the bulk, non-Hermitian skin effects, or trivial boundary contributions~\cite{Qi_Topological_2011,chu2023broad,allen_spatially_2016,aharon2021long,zhu2017edge}. However, the extracted widths are upper bounds on the intrinsic width of the localized channels~\cite{hui2014proximity}, because the Fourier-based reconstruction is limited by the finite magnetic-field range of the measured interference pattern. As shown in Supplementary Fig. S7, even for an ideal SQUID interference pattern, the extracted supercurrent peaks have finite widths. Thus, the reconstructed supercurrent density profile reliably identifies where the supercurrent predominantly flows, but does not by itself provide a precise quantitative measure of the intrinsic width or magnitude of the supercurrent peaks~\cite{hui2014proximity}.

\begin{figure}[t]
\centering
\includegraphics[width=\textwidth]{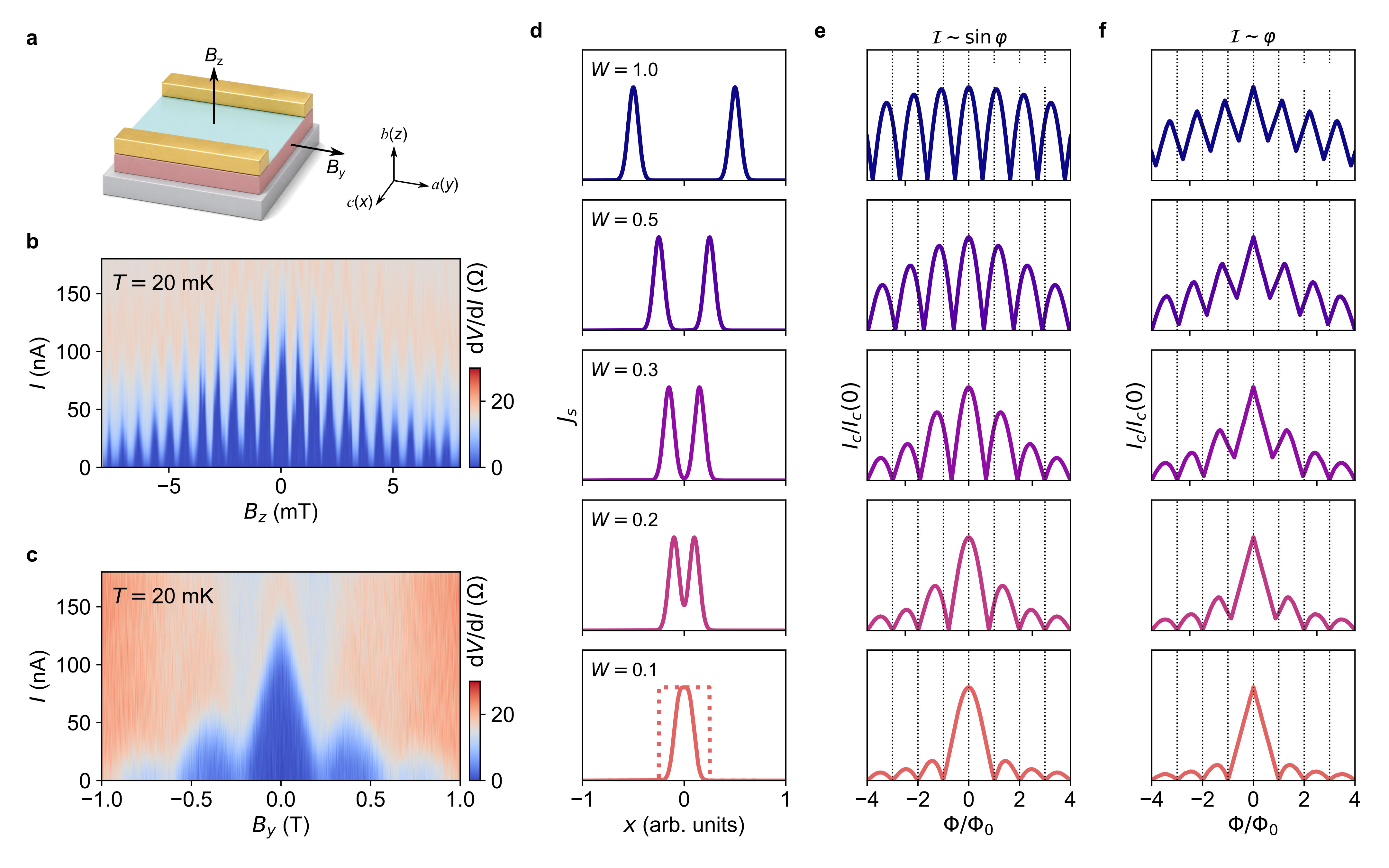}
\caption{\textbf{Field-orientation dependence of facet-selective supercurrent.} \textbf{a}, Schematic of the NbN/ZrTe$_5$ Josephson junction (Dev~3), where the current flows along the \textit{c}-axis. \textbf{b}, Superconducting interference patterns measured at 20~mK for magnetic field  $B_z$ applied along the \textit{b}-axis, showing behavior similar to that in Fig.~\ref{Fig-edgeSC}. \textbf{c}, Same measurement performed for magnetic field $B_y$ applied along the \textit{a}-axis. \textbf{d}, The supercurrent density profiles for the parametrized junction width $W = 1, 0.5, 0.3, 0.2,$ and $0.1$. \textbf{e,f}, Corresponding simulated interference patterns for a sinusoidal (\textbf{e}) and a sawtooth (\textbf{f}) current–phase relation (see Supplementary Note 7 for details). For reference, a uniform supercurrent profile (indicated by the dotted box in the bottom panel of \textbf{d}) produces the same interference pattern as a centrally peaked supercurrent profile.}
\label{Fig-JJ_angle}
\end{figure}

Although the observed surface-dominated supercurrent can in principle arise from trivial boundary channels due to local doping or band bending near the edges, our results are inconsistent with this picture based on the following three observations. First, SQUID-like patterns appear for two separate Al/ZrTe$_5$ devices. Moreover, a NbN/ZrTe$_5$ junction oriented along the $c$-axis produces SQUID-like oscillations (Fig.~\ref{Fig-JJ_angle}\textbf{a,b}), demonstrating reproducibility across different superconductors and crystallographic orientations. Second, the superconducting interference pattern of the NbN/ZrTe$_5$ junction for magnetic field $B_y$ (perpendicular to the $b$--$c$ plane) shows a strongly enhanced central lobe and rapidly decreasing side-lobe amplitudes (Fig.~\ref{Fig-JJ_angle}\textbf{c}), reflecting a more homogeneous current distribution on the $b$-$c$ facet. The distinct difference between the interference patterns for $B_y$ and $B_z$ is a direct consequence of the facet-selectivity of the WTI, in which supercurrent is mediated by  conducting boundary states that appear only on the $a$-$b$ and $b$-$c$ facets. Such a field-orientation contrast is difficult to reconcile with a trivial mechanism. Third, the ballistic nature of the supercurrent, as evidenced by the exponential temperature dependence of $I_c$ (Fig.~\ref{Fig-inducedSC}\textbf{d}), is incompatible with a trivial diffusive transport scenario.

\subsection*{Non-sinusoidal current–phase relation: ballistic topological channels}

Further evidence for ballistic transport emerges from the shape of the interference lobes, which depends strongly on the current–phase relation (CPR) of the junction. For diffusive junctions and low-transparency ballistic junctions, the CPR is essentially sinusoidal~\cite{golubov2004current}, i.e., $\mathcal{I} \sim \sin\varphi$, and the simulated interference patterns exhibit rounded lobes irrespective of the supercurrent density profile $J_s(x)$ (Fig.~\ref{Fig-JJ_angle}\textbf{d,e}). In contrast, in the long ballistic limit, we expect a sawtooth CPR~\cite{golubov2004current}, i.e., $\mathcal{I} \sim \varphi$ for $|\varphi| < \pi$, which leads to triangular lobes (Fig.~\ref{Fig-JJ_angle}\textbf{f}). The sharp lobes observed in Fig.~\ref{Fig-JJ_angle}\textbf{b,c} indicate that the supercurrent is carried by ballistic channels.

The interference pattern for $B_z$ qualitatively resembles the simulated profiles for $W =$~0.5-1.0, for which the surface supercurrent peaks in $J_{\mathrm{s}}(x)$ are localized at the two junction edges. When an in-plane field is applied to the same junction, the area threaded by $B_y$ lies on the conducting $b$-$c$ facets, and a homogeneous supercurrent producing a pattern like that shown in the bottom panel of Fig.~\ref{Fig-JJ_angle}\textbf{f} is therefore expected. However, the experimentally observed interference pattern (Fig.~\ref{Fig-JJ_angle}\textbf{c}) is similar to the simulations for $W =$~0.2-0.3, which corresponds to two edge‑localized supercurrent channels that are partially merged. While this scenario naturally explains the near flux‑periodic modulation of the critical current (Supplementary Note 5), the appearance of edge channels is puzzling. A plausible microscopic origin of these conducting edges is the emergence of higher‑order topological hinge states at the intersection of the $b$-$c$ and $c$-$a$ facets~\cite{schindler2018higher_sci,luo2021higher}. However, on the basis of superconducting quantum interference measurements alone, we cannot unambiguously distinguish between hinge modes and surface states.

\section*{Discussion}

In this work, we identify four experimental signatures of facet‑selective ballistic supercurrent: (i) SQUID‑like superconducting interference patterns, which reveal supercurrent peaks localized near two spatially separated facets, (ii) distinct interference patterns for two different field orientations reflecting facet selectivity, (iii) an exponential temperature dependence of the critical current consistent with ballistic transmission in long junctions, and (iv) sharp interference lobes indicative of a sawtooth current–phase relation in the ballistic regime. Together, these observations support a self-consistent picture of ballistic supercurrent mediated by WTI boundary states in ZrTe$_5$. The reproducibility across two superconductors (Al and NbN), two junction orientations (along the $a$- and $c$-axes), and three independent devices suggests that our results originate from intrinsic properties of ZrTe$_5$, rather than from device-specific artifacts. More broadly, we demonstrate that bulk topology can be read out directly from the spatial distribution of proximity-induced supercurrent, and we identify weak topological insulators as a suitable material system for facet-resolved superconductivity. 


\section*{Methods}\label{sec11}
\textit{Fabrication.}
The samples were prepared by mechanically exfoliating ZrTe$_5$ crystals onto SiO$_2$/Si substrates using the Scotch-tape method inside a glovebox, followed by immediate encapsulation with a resist layer to prevent degradation. Subsequently, the flakes were patterned into well-defined Hall-bar geometries using electron-beam lithography (EBL) and Ar$^{+}$ ion milling. Ti (15~nm)/Au (250~nm) contacts were then deposited by electron-beam evaporation on the patterned flakes using a second EBL step. To fabricate the Josephson junctions, a single EBL step was used to define the superconducting leads, followed by the deposition of Pd (4~nm)/Al (200~nm) via electron-beam evaporation or Pd (4~nm)/NbN (120~nm) via dc sputtering. Both types of devices were capped with 15 nm thick Al$_2$O$_3$ using e-beam evaporation to prevent sample degradation. The dimensions of the devices are listed in Supplementary Table S1.\\

\noindent \textit{Transport measurements.}  
Electronic transport measurements were carried out using a lock-in amplifier (SR830) integrated with a Quantum Design cryogen-free PPMS DynaCool system. Supercurrent measurements were performed in a Bluefors dilution refrigerator with a base temperature of 15~mK, employing a combined DC and AC current bias technique. Magnetic fields were applied using a three-axis vector magnet, enabling measurements for both in-plane and out-of-plane orientations.\\

\noindent \textit{STM measurements.}
Scanning tunnelling microscopy and spectroscopy experiments were performed in a home-built STM setup operated under ultrahigh vacuum conditions at $T$ = 5.1 K. Before measurements, the tip was conditioned on an Au(111) surface prepared by sputtering and annealing cycles. Spectroscopic measurements were acquired using a lock-in technique. The scan parameters for the data presented in Fig.~\ref{Fig-transport}\textbf{e} were $U = -1$~V and $I$ = 100 pA, while those in Fig.~\ref{Fig-transport}\textbf{g} were $U = -190$~mV, $I$ = 200 pA, and $U_{mod}$ = 1 mV.\\

\bmhead{Acknowledgements}

The authors acknowledge financial support from KAW-WISE (Wallenberg Initiative Materials Science for Sustainability), Swedish Research Council VR project grants (No. 2025-03702), 2D-TECH VINNOVA competence center (No. 2019-00068), Army Research Office (ARO) (Grant No. W911NF2210053), FLAG-ERA project MagicTune (No. 2023-06210), Areas of Advance Nano and Materials science at Chalmers University of Technology, and Nanofabrication laboratory MyFab at Chalmers University of Technology. P.R. acknowledges helpful discussions with A. Udupa and technical assistance from N. Trnjanin.

\bmhead{Author Contributions}
P.R. and S.P.D. conceived the idea and coordinated the project. P.R., A.K. and L.P. fabricated the devices. P.R. and A.K. performed the supercurrent measurements. P.R., A.K. and L.P. performed the Hall bar measurements. P.S. and X.H. performed  the STM experiments and analyzed the data. P.R. and A.K. analyzed the transport data. P.R. and A.K. wrote the manuscript with input from P.S., X.H., I.C., S.G., M.B., J.Å., F.L., T.B. and S.P.D.

\bibliography{sn-bibliography}

\end{document}